\def\pds{\kern+0.1em /\kern-0.55em \partial}
\def\lts#1{\kern+0.1em /\kern-0.65em #1}
\def\half{\frac{1}{2}}

\def\bey{\begin{eqnarray}}
\def\eey{\end{eqnarray}}
\def\be{\begin{equation}}
\def\ee{\end{equation}}
\def\ba{\begin{array}}
\def\ea{\end{array}}
\def\gm{\gamma}
\def\Gm{\Gamma}

\def\ld{\lambda}

\def\af{\alpha}
\def\sg{\sigma}
\def\Sg{\Sigma}

\def\om{\omega}
\def\r{\rho}

\def\bt{\beta}
\def\vep{\varepsilon}
\def\ep{\epsilon}

\def\Dt{\Delta}

\def\pp{\partial}
\def\pp{\partial}

\def\ra{\rightarrow}
\def\nnb{\nonumber}
\documentclass[onecolumn,showpacs,preprintnumbers,amsmath,amssymb]{revtex4}
\usepackage{graphicx}
\usepackage{fancyhdr}
\usepackage{dcolumn}
\usepackage{bm}
\begin{document}
\preprint{ }

\title{ Effects of medium-induced $\rho-\omega$ meson mixing on
the equation of state in isospin-asymmetric nuclear matter}
\author{Wei-Zhou Jiang$^{1,2,3}$  and  Bao-An Li$^1$ }
\affiliation{$^1$ Department of Physics and Astronomy, Texas A\&M
University-Commerce, Commerce, TX 75429, USA } \affiliation{ $^2$
Department of Physics, Southeast University, Nanjing 211189,
China} \affiliation{ $^3$ Institute of Applied Physics,
  Chinese Academy of Sciences, Shanghai 201800, China}
\date{}

\begin{abstract}
\baselineskip18pt We reexamine effects of the $\rho-\omega$ meson mixing
mediated by nucleon polarizations on the symmetry energy in
isospin-asymmetric nuclear matter. Taking into account the rearrangement term
neglected in previous studies by others, we evaluate the $\rho-\omega$ mixing
angle in a novel way within the Relativistic Mean-Field  Models with and
without chiral limits. It is found that the symmetry energy is significantly
softened at high densities contrary to the finding in earlier studies.  As
the first step of going beyond the lowest-order calculations, we also solve
the RPA equation for the $\rho-\omega$ mixing. In this case, it is found that
the symmetry energy is not only significantly softened by the $\rho-\omega$
mixing at supra-saturation densities, similar to the lowest-order
$\rho-\omega$ mixing, but interestingly also softened at subsaturation
densities. In addition, the softening of the symmetry energy at subsaturation
densities can be partly suppressed by the nonlinear self-interaction of the
$\sg$ meson.
\end{abstract}
\baselineskip 18.6pt 
  \maketitle

\baselineskip 20.6pt
\section{Introduction}
Many hadronic and partonic approaches have been used in studying the
charge symmetry breaking (CSB) and its effect on few-body and
bulk-matter observables, see, e.g., refs.~\cite{mi90,mac} for
reviews. The CSB effect can be displayed in the charge-conjugate
systems such as the proton-proton and neutron-neutron binary systems
whose scattering lengths in the $^1S_0$ state: $a_{nn}$ and $a_{pp}$
differ by 10\% after deducting the electromagnetic interaction. The
CSB can also be used to explain the well-known Nolen-Schiffer anomaly
of light mirror nuclei. Among the hadronic approaches, the CSB effect
has been incorporated into the many-body theories by employing
explicitly charge-dependent nucleon-nucleon interactions, e.g., the
Bonn\cite{ma01}, Reid93\cite{bord} and V18 potentials~\cite{wi95}.
These interactions have been adjusted to
reproduce the free-space nucleon-nucleon scattering data. However,
the CSB effects in free-space and/or symmetric nuclear matter are
normally very small. For example, the CSB-induced effects in
symmetric nuclear matter with the charge-dependent Bonn potential
were shown to be quite small ~\cite{ma01}. The results with the
charge-dependent Reid93 potential also showed that the CSB effect on
the equation of state (EOS) even in isospin-asymmetric nuclear matter
is negligible~\cite{bord}.

To explore the possible CSB mechanisms at the hadronic level, the
$\r^0-\om$ meson mixing (ROM)~\cite{coon,mc,pi93,ki96,du97,du00}
and the hadron mass splitting effects~\cite{ma01,gqli98} have also
been studied extensively. Within the meson mixing picture, the two
charge neutral vector mesons undertake a transition to each other
through a baryon-antibaryon polarization. An essential point is
the incomplete cancellation between the opposite contributions
from proton  and neutron loops to the transition
amplitude~\cite{pi93}. In nuclear medium, the polarization
acquires an overwhelming enhancement through the particle-hole
excitations. Very interestingly, the ROM gets significantly
amplified in isospin-asymmetric nuclear matter, see, e.g.,
refs.~\cite{du97,ki97,is00}, because of the different numbers of
neutrons and protons available. In a nutshell, the different
stacking of protons and neutrons in the Fermi sea becomes a
naissant factor to intrigue the CSB in isospin-asymmetric nuclear
matter due to the medium-induced ROM. Loosely, we may call this
the in-medium CSB.

The ROM in asymmetric nuclear matter also results in significant
modifications to properties of the isovector meson $\rho$ and its
couplings with nucleons~\cite{ki96,du97,du00,ki97,is00}. Since the
potential part of the nuclear symmetry energy is dictated by the
exchange of $\rho$ mesons, at least within the Relativistic Mean-Field
(RMF) models, corresponding modifications to the density dependence of
the symmetry energy are thus expected. While most earlier
studies~\cite{du97,du00,mo03,ag04,mu07} have focused on investigating
the in-medium meson spectra, a few studies ~\cite{ki97,is00} have
indeed looked at effects of the medium-induced ROM on the symmetry
energy. It was found that the $\r-\om$ mixing angle reaches its maximum
of about 45$^\circ$ in isospin-asymmetric nuclear matter. Moreover, the
symmetry energy was sharply stiffened at supra-normal densities.
However, the rearrangement term, which is crucial for the thermodynamic
consistency in deriving the pressure, was neglected in these  RMF
studies~\cite{ki97,is00}. We find that the rearrangement term actually
causes the pressure to exceed the energy density when the mixing angle
$\theta$ is close to 45$^\circ$. This certainly violates the causality,
namely, the work done by the pressure of the system should be less than
its total energy. In this work, using the RMF Lagrangian constructed by
us recently~\cite{ji07,jia07} including constraints of a mass dropping
scenario according to the Brown-Rho (BR)
scaling~\cite{br91,br07,li95,ha03,ha07}, we calculate the $\r-\om$
mixing angle in asymmetric matter in a novel way. We then examine
effects of the medium-induced ROM on the in-medium masses of the $\r$
and $\om$ mesons and the density dependence of the symmetry energy. As
the first step of going beyond the lowest-order nucleon polarization,
we also evaluate the $\rho-\omega$ mixing within the random phase
approximation (RPA). Results from the two different levels of
approximations will be compared with each other. Moreover, in order to
examine the model dependence we also perform the calculations with the
nonlinear RMF model.

The paper is organized as follows. In Section \ref{ROM}, we
introduce the medium-induced ROM within the RMF model and the novel
way to obtain the $\r-\om$ mixing angle. We then summarize briefly
the formalisms for the ROM within both the lowest-order nucleon
polarization and the RPA. Results and discussions are presented in
Section \ref{result}. A summary is given in Section \ref{summary}.

\section{Medium-induced $\rho-\omega$ meson mixing within the RMF model}
\label{ROM}

The following lagrangian with the chiral limit serves as the starting
point~\cite{ji07,jia07}:
\begin{eqnarray}
 {\mathcal L}&=&
{\overline\psi}[i\gm_{\mu}\partial^{\mu}-M +g^*_{\sg}\sg-G^*_{\om }
\gm_{\mu}V_\om^{\mu}-G^*_{\r}\gm_\mu \tau_3 V_\r^\mu]\psi\nnb\\
&&
+\frac{1}{2}(\partial_{\mu}\sg\partial^{\mu}\sg-m_{\sg}^{*2}\sg^{2})
 - \frac{1}{4}F_{\mu\nu}F^{\mu\nu}+
      \frac{1}{2}m_0^{*2}{V_\om}_{\mu}V_\om^{\mu}\nnb\\
&&
       - \frac{1}{4}B_{\mu\nu} B^{\mu\nu}+
\frac{1}{2}m_1^{*2} V_{\r\mu} V_\r^{\mu}+\epsilon
V_{\om\mu}V_{\r}^{\mu}\label{eq:lag1}
\end{eqnarray}
where $\psi,\sigma,V_\om$, and $V_\r$ are the fields of the nucleon,
scalar, vector, and isovector-vector mesons, with their masses $M,
m^*_\sg,m^*_0$, and $m^*_1$, respectively. The meson coupling
constants and masses with asterisks denote the density dependence,
given by the BR scaling with well restrained forms and
parameters~\cite{ji07,jia07}. We note that the BR scaling is still a
phenomenological ansatz to mimic the properties of the partial
restoration of the chiral symmetry although QCD-based effective
field theories and models support the mass dropping scenario.  For
instance,  as an effective QCD field theory, the hidden local
symmetry theory has been developed to include the $\r$ meson in
addition to the pion in the framework of the chiral perturbative
theory by Harada and Yamawaki and it is shown that  the $\rho$ meson
becomes massless at the chiral limit~\cite{ha03,ha07}. Moreover, the
controversy of the BR scaling is still unsettled by recent
experiments~\cite{br07}, also see the discussions in
Ref.~\cite{ji07}. Here we resort to the mass dropping scenario
according to the BR scaling mainly due to its simplicity.

The term $\epsilon V_{\om\mu}V_{\r}^{\mu}$ in the lagrangian that mixes
the isoscalar and isovector mesons breaks the  charge symmetry in the
isospin space. In the free space, the parameter $\epsilon$ represents
the amplitude of the fundamental CSB within the ROM picture. In
isospin-asymmetric nuclear matter, since the medium effect is
dynamically induced, it can be conveniently expressed as
\begin{equation}\label{ems}
    \epsilon=m_0^* m_1^*\gm,
\end{equation}
in terms of a dimensionless parameter $\gm$. The symmetry breaking term
can be exorcized at the cost of introducing a mixing angle with the
unitary transformation~\cite{ki96,bi99}:
\begin{equation}
\label{trans} V_\om=b_0\sin\theta + \om \cos\theta, { }\ V_\r=b_0
\cos\theta -\om\sin\theta,
\end{equation}
with
\begin{equation}\label{eq:eps}
 \tan 2\theta=\frac{2\epsilon}{{m_1^*}^2-{m_0^*}^2}.
\end{equation}
The lagrangian thus becomes
\begin{eqnarray} {\mathcal L}&=&
{\overline\psi}[i\gamma_{\mu}\partial^{\mu}-M
+g^*_{\sigma}\sigma-g^*_{\omega }
\gamma_{\mu}\omega^{\mu}-g^*_{\rho}\gamma_\mu \tau_3 b_0^\mu
]\psi\nnb\\
&& +\frac{1}{2}(\partial_{\mu}\sigma\partial^{
\mu}\sigma-m_{\sigma}^{*2}\sigma^{2})
- \frac{1}{4}F_{\mu\nu}F^{\mu\nu}+ \frac{1}{2}m_{\omega}^{*2}\omega_{\mu}%
\omega^{\mu}\nonumber \\
&&  - \frac{1}{4}B_{\mu\nu} B^{\mu\nu}+ \frac{1}{2}m_{\rho}^{*2}
b_{0\mu} b_0^{\mu}, \label{eq:lag2}
\end{eqnarray}
where both $g^*_\om$ and $g^*_\r$ couple differently to the proton
and neutron~\cite{ki96}:
\begin{equation}\label{eq:coup1}
  g_{{\rm p}\om}^* =G_\om^*\cos\theta-G_\r^*\sin\theta,\hbox{ } g_{{\rm n}\om}^*=
  G_\om^*\cos\theta+G_\r^*\sin\theta
\end{equation}
\begin{equation}\label{eq:coup2}
 g_{{\rm p}\r}^* = G_\r^*\cos\theta+G_\om^*\sin\theta, \hbox{ } g_{{\rm n}\r}^*=
G_\r^*\cos\theta-G_\om^*\sin\theta
\end{equation}
The $\om$ and $\r$ masses are given by
\begin{eqnarray}\label{eq:mm}
{m_\om^*}^2 &=& \frac{{m_0^*}^2+{m_1^*}^2}{2}+
\frac{{m_0^*}^2-{m_1^*}^2}{2\cos2\theta}, \nnb\\
{m_\r^*}^2 &=& \frac{{m_0^*}^2+{m_1^*}^2}{2}-
\frac{{m_0^*}^2-{m_1^*}^2}{2\cos2\theta},
\end{eqnarray}
where the masses of $\r$ and $\om$ mesons  shift oppositely with
$\theta$, required by the transformational unitarity in
Eq.(\ref{trans}).  The energy density, based on Eq.(\ref{eq:lag2}),
reads,
 \bey
  \label{eqe1}
 {\mathcal{E}}&=&\frac{(g_{{\rm p}\om}^*\r_p+g_{{\rm n}\om}^*\r_n)^2}{
2{m_\om^*}^2} +\frac{(g_{{\rm p}\r}^*\r_p-g_{{\rm n}\r}^*\r_n)^2}{
2{m_\r^*}^2}\nnb\\
&&+ \half {m_\sg^*}^2\sg^2 +\sum_{i=p,n}
 \frac{2}{(2\pi)^3}\int_{0}^{{k_F}_i}\! d^3\!k~ E^*_i,
 \eey
where the $\theta$ is embedded in both meson masses and their
couplings to nucleons.  Given the $\r-\om$ meson mixing by the
nucleon polarization that is beyond the mean field, the rearrangement
term is inevitably induced to keep the Lorentz invariance of the RMF
model. This is a general case in the density-dependent RMF models
where the density dependence is incorporated beyond the mean field,
e.g., see~\cite{ji07,jia07,len95}. Besides the contribution from the
BR scaling, the density-dependent mixing angle results in a new
source of the rearrangement term in the pressure
 \bey\label{eqp1}
 p&=&\frac{(g_{{\rm p}\om}^*\r_p+g_{{\rm n}\om}^*\r_n)^2}{
2{m_\om^*}^2} +\frac{(g_{{\rm p}\r}^*\r_p-g_{{\rm n}\r}^*\r_n)^2}{
2{m_\r^*}^2}\nnb\\
& &- \half {m_\sg^*}^2\sg^2 -\Sg_0\r+
 \frac{1}{3}\sum_{i=p,n}\frac{2}{(2\pi)^3}\int_{0}^{{k_F}_i}\! d^3\!k
 ~\frac{{\bf k}^2}{E^*_i}.
 \eey
Here, an additional $\theta$-related part $-\pp {\mathcal
E}/\pp\theta\cdot\pp \theta/\pp\r$ neglected in
Refs.~\cite{ki97,is00} makes the rearrangement term $\Sg_0$ more
involved than that in Ref.~\cite{ji07}, and the final result is
obtained numerically.

\begin{figure}[tbh]
\begin{center}
\vspace*{-20mm}\includegraphics[width=0.7\textwidth]{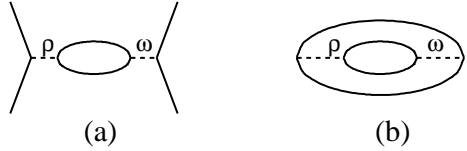}
\vspace*{-20mm}
\end{center}
\caption{Diagrams for (a) the scattering, and (b) the corresponding
contribution to the potential energy density. \label{fmix1}}
\end{figure}

A key point is to derive the $\r-\om$ mixing angle through the nucleon
polarization. In principle, one can firstly determine the ROM parameter
$\epsilon$ or $\gm$  directly from the polarization with the relation
$g_{\mu\nu}\epsilon=\Pi^{\r\om}_{\mu\nu}(q^2)$ according to the
lagrangian (\ref{eq:lag1}), similar to the case in the free
space~\cite{pi93}, and then the mixing angle can be obtained through
Eq.(\ref{eq:eps}). However, the ROM parameters $\epsilon$ and $\gm$
obtained in this way have a complicated structure in momentum
space~\cite{pi93}. In order to obtain momentum-independent ROM
parameters in the mean-field approximation, an average procedure is
necessary. As mentioned in the Introduction, the mixing angle obtained
with such an averaging procedure in Refs.~\cite{ki97,is00} is large
enough to cause the violation of causality.  Here, we pursue a new
approach. We first evaluate the total energy density from the
lagrangian (\ref{eq:lag1}) but with the explicit ROM replaced by the
polarization diagram in bulk matter, see Fig.~\ref{fmix1}. Since the
polarization diagram is a contribution beyond the RMF approach, the
polarization in Fig.~\ref{fmix1} is not limited only to its temporal
component considered in Refs.~\cite{ki97,is00}. The ROM angle appearing
in the lagrangian (\ref{eq:lag2}) is then determined by reproducing
this total energy density using Eq.(\ref{eqe1}). Meanwhile, the ROM
parameter $\ep$ or $\gm$ can be calculated through Eq.(\ref{eq:eps}).
Nevertheless, here we indeed evaluate the ROM parameter $\ep$ in the
lagrangian (\ref{eq:lag1}) through the $\r^0-\om$ transition amplitude.
In isospin-asymmetric matter, the medium effect is dynamically induced
due to the incomplete cancellation of the proton and neutron loops. In
the lowest order of the meson-nucleon coupling, this is equivalent to
computing the contribution of the lowest-order nucleon polarizations to
the energy density. To include the high-order effect, we need to solve
the RPA (Dyson) equation to sum up relevant nucleon loops to all
orders.

The energy density can be written quite generally
as~\cite{se87,bu87,br90} ${\mathcal E}={\mathcal T}+{\mathcal V}/2$,
where the ${\mathcal T}$ and ${\mathcal V}$ are the kinetic and
potential parts, respectively. The polarization diagram gives rise
to a term ${\mathcal V}^\prime$ adding to the potential energy
density ${\mathcal V}$. Considering ${\mathcal V}^\prime$ is just a
few percents compared to the vector potential even in pure neutron
matter, we neglect its direct  correction to the meson fields (or,
nucleon self-energies) that are solved in the RMF equations. In
fact, a full calculation with the meson fields including this term
brings about the momentum dependence of the nucleon potential which
should be fulfilled at least in the Hartree-Fock (HF) framework.
This is, however, beyond the present RMF treatment. In the
evaluation of the ${\mathcal V}^\prime$, the polarization tensor is
wrapped up by the nucleon current  only in the Fermi sea. This
approach is actually widely used in the study of bulk
matter~\cite{se87,bu87,br90}. The ${\mathcal V}^\prime$ is thus
written as
\begin{eqnarray}\label{edpot1}
{\mathcal V}^\prime &=& -{G_\om^*} {G_\r^*}\int
   \frac{d^3\!p_1}{(2\pi)^3}\frac{d^3\!p_2}{(2\pi)^3} {\rm Tr}
[\frac{\lts{p}_1^*+M^*_i}{2E^*_{p_1i}}\gm^\mu\tau_3
\frac{\lts{p}_2^*+M^*_i}{2E^*_{p_2i}}\gm^\nu]\nnb \\
 && \times\frac{\Pi^{\rho\om}_{\mu\nu}(q)}{({m_0^*}^2-q^2)({m_1^*}^2-q^2)},
\end{eqnarray}
where $p_i^{*0}=E_{pi}^*=\sqrt{{\bf p}^2+{M_i^*}^2}$,  ${\bf q}={\bf
p}_2-{\bf p}_1$ and $q_0=E^*_{p2}-E^*_{p1}$. The polarization tensor is
given by
\begin{equation}\label{eq:pola2}
\Pi^{\rho\om}_{\mu\nu}(q) =-i{G_\om^*} {G_\r^*}\int\frac{d^4\!k}{(2\pi)^4}
{\rm Tr}
  [\gm_\mu G(k)\gm_\nu\tau_3 G(k+q)],
  \end{equation}
where $G(k)$ is the nucleon propagator, and the trace is over both the
isospin ($\tau$) and spin ($\gm$) matrices. Here, the polarization
tensor is given in the lowest order, while the high-order corrections
will be considered later on by solving the RPA (Dyson) equation. As
seen in Eq.(\ref{edpot1}), only the real part of the polarization is
needed to evaluate the ${\mathcal V}^\prime$. Usually, the imaginary
part concerns the stability of the mean field in small oscillations.
Indeed, it also comes into play in the real part of the polarization by
taking into account the high-order corrections in the RPA equation.
First of all, we consider the case with the lowest-order polarization.
By choosing the frame $(q_0, \bar{q}, 0, 0)$ with
$q^2=q_0^2-\bar{q}^2$, one can define the longitudinal ($L$) and
transverse ($T$) components~\cite{ch77}:
\begin{equation}\label{eq:pola1}
\Pi_L=\Pi_{33}-\Pi_{00},\hbox{ } \Pi_T=\Pi_{11}=\Pi_{22}.
\end{equation}
In free space $\Pi_L=\Pi_T=\Pi^F(q^2)$. Here, the tensor coupling
constant of $\r$ meson is not included because it is in principle
nonrenormalizable. In the medium, the polarization tensor is usually
decomposed into Feynman and density-dependent parts:
$\Pi=\Pi^F+\Pi^D$, while the former is just the one in free space
but with the in-medium nucleon mass. The trace over the isospin
makes the Feynman parts of the proton and neutron polarizations to
cancel each other, leading to~\cite{pi93}
\begin{equation}\label{eq:pola3}
\Pi^{\rho\om,F}(q^2) =\frac{q^2}{2\pi^2}{G_\om^*} {G_\r^*}\int_0^1d\!x
x(1-x)\ln\frac{{M^*_p}^2-x(1-x)q^2}{{M^*_n}^2-x(1-x)q^2}.
\end{equation}
In the medium, the longitudinal and transverse components of the
density-dependent part are not equal~\cite{ch77}:
\begin{eqnarray}
  \Pi^{\rho\om,D}_L &=&-8q^2{G_\om^*} {G_\r^*}\sum_{i=p,n}s\int
  \frac{d^3\!k}{(2\pi)^3} \frac{\Theta(k_{F_i}-|k|)}{E_{ki}^*}
  \frac{{E_{ki}^*}^2-{\bf k}^2\chi^2}{(q^2)^2-4(k\cdot q)^2}\nnb \\
  \Pi^{\rho\om,D}_T &=&-8{G_\om^*} {G_\r^*}\sum_{i=p,n}s\int
  \frac{d^3\!k}{(2\pi)^3} \frac{\Theta(k_{F_i}-|k|)}{E_{ ki}^*}
  \frac{q^2{\bf k}^2(\chi^2-1)/2+(k\cdot q)^2}
  {(q^2)^2-4(k\cdot q)^2},
\end{eqnarray}
where  $\Theta$ is the step function, $k_{F_i}$ with $i=p,n$ are
respectively the proton and neutron Fermi momenta, $k\cdot
q=E_{ki}^*q_0-|{\bf k}|\bar{q}\chi$, $\chi=\cos\angle{\bf k\cdot q}$,
and  $s=\pm1$ for proton and neutron, respectively. Note that only the
real parts of the polarization are given here. With the chosen frame
for $q_\mu$, the polarization can be written as:
\begin{eqnarray}\label{eq:polam}
\Pi^D_{\mu\nu} &=&(-g_{\mu\nu}+q_\mu q_\nu/q^2)\Pi_L^D, \hbox { }
\mu,\nu=3,0\nnb \\
\Pi^D_{\mu\nu} &=&-g_{\mu\nu}\Pi_T^D, \hbox { } \mu,\nu=1,2.
\end{eqnarray}
Substituting Eqs.(\ref{eq:pola3}),(\ref{eq:polam}) in (\ref{edpot1})
and considering the conservation of baryon number, we obtain
\begin{eqnarray}\label{edpot2}
{\mathcal V}^\prime &=& \frac{4{G_\om^*} {G_\r^*}}{(2\pi)^4}
\sum_{i=p,n}s\int_0^{k_{Fi}} p_1^2 p_2^2 d\!p_1 d\!p_2 \int_{-1}^{1} d\!\chi_p
\left\{\frac{p_1p_2\chi_p+{M^*_i}^2}{E^*_{p_1i}E^*_{p_2i}}\right.
\Pi^{\rho\om}_L(q) \nnb \\
&&\left.+(\frac{{M^*_i}^2}{E^*_{p_1i}E^*_{p_2i}}
-1)\Pi^{\rho\om}_T(q)\right\}  \frac{1}{({m_0^*}^2-q^2)({m_1^*}^2-q^2)},
\end{eqnarray}
with $\chi_p=\cos\angle {\bf p}_1\cdot {\bf p}_2$. With this
expression that surrogates the explicit ROM, we can obtain
straightforwardly the total energy density from the lagrangian
(\ref{eq:lag1}) in the RMF approximation.

In the above, we calculated the additional energy density from the
lowest-order polarization of the ROM without including the high-order
ones. A more elaborate treatment needs to include the ROM and the
high-order corrections altogether by solving the RPA equation. The RPA
equation for the polarizations is given as:
\begin{equation}\label{rpa1}
    \tilde{\Pi}_{\mu\nu}=\Pi_{\mu\nu}+\tilde{\Pi}_{\mu\ld}{\mathcal
    D}^{\ld\tau}\Pi_{\tau\nu}.
\end{equation}
This equation can be decomposed into the longitudinal and transverse parts as
\begin{eqnarray}\label{rpa2}
  \tilde{\Pi}_T &=& \Pi_T+\tilde{\Pi}_T{\mathcal D}_T\Pi_T, \nnb\\
 \tilde{\Pi}_L &=& \Pi_L+\tilde{\Pi}_L{\mathcal D}_L\Pi_L,
\end{eqnarray}
where the lowest-order polarization matrices are given as:
\begin{equation}\label{rpa3}
    \Pi_T=\left(\begin{array}{cc}
  \Pi^{\om\om}_T & \Pi^{\rho\om}_T \\
  \Pi^{\rho\om}_T & \Pi^{\rho\rho}_T \\
\end{array}\right),~
\Pi_L=\left(
\begin{array}{ccc}
  \Pi^{\sg\sg} & \Pi^{\sg\om}_0 & \Pi^{\sg\rho}_0 \\
  \Pi^{\sg\om}_0 & \Pi_{00}^{\om\om} & \Pi_{00}^{\rho\om} \\
  \Pi^{\sg\rho}_0 & \Pi_{00}^{\rho\om} & \Pi_{00}^{\rho\rho} \\
\end{array}
\right),
\end{equation}
and the meson propagator matrices are diagonal~\cite{ho91,re99}
\begin{equation}\label{prop1}
    {\mathcal D}_T=\left(
\begin{array}{cc}
  D_\om & 0 \\
  0 & D_\rho \\
\end{array}\right),~{\mathcal D}_L=\left(
\begin{array}{ccc}
  \Delta & 0  & 0  \\
   0 &q^2 D_\om /\bar{q}^2 & 0  \\
  0  & 0  &q^2 D_\rho /\bar{q}^2  \\
\end{array}%
\right),
\end{equation}
with $\Delta=1/(q^2-m_\sg^{*2}+i\vep)$, and
$D_{\om,\rho}=1/(q^2-m_{0,1}^{*2}+i\vep)$. Now, the polarizations of
the ROM in the RPA can be written explicitly as:
\begin{equation}\label{rpa4}
\tilde{\Pi}_L^{\rho\om}=\frac{q^2}{\bar{q}^2}\tilde{\Pi}^{\rho\om}_{00}=
\frac{{\Pi}_L^{\rho\om}+\Delta(\frac{q^2}{\bar{q}^2}
\Pi_0^{\sg\om}\Pi_0^{\sg\rho}-\Pi^{\sg\sg}\Pi_L^{\rho\om})}{\epsilon_L},
~\tilde{\Pi}_T^{\rho\om}=\frac{{\Pi}_T^{\rho\om}}{\epsilon_T},
\end{equation}
where the dielectric functions $\epsilon_L$ and $\epsilon_T$ are the
determinants of the matrices $(1-{\mathcal D}_L\Pi_L)$ and
$(1-{\mathcal D}_T\Pi_T)$, respectively. Their explicit expressions,
together with definitions of the polarizations in matrices
(\ref{rpa3}), are given in appendix A. Note that the imaginary parts of
the polarizations are also included in the RPA equation, and thus
finally we take the real part of the $\tilde{\Pi}^{\rho\om}$ in the
calculation of the energy density. Here, except for ${\Pi}^{\rho\om}$,
the vacuum part is neglected for all other polarizations since we work
in the RMF approximation. Substituting above expressions into
Eq.(\ref{edpot2}), we may evaluate the additional energy density within
the RPA.

\section{Results and discussions}
\label{result}

Our numerical calculations are carried out based on the
SLC~\cite{jia07} and the well-known NL3~\cite{nl3} parameter sets. The
SLC parameter set was obtained by reproducing the pressure profile
constrained at supra-saturation densities by the collective flow data
from relativistic heavy-ion reactions~\cite{da02} and the density
dependence of the symmetry energy at subsaturation densities obtained
from studying isospin diffusion in heavy-ion reactions at intermediate
energies~\cite{ts04,ch05,li05}, besides the saturation properties of
nuclear matter. In the following, we first carry out numerical
calculations within the lowest-order ROM picture. This facilitates easy
and direct comparisons with earlier studies by others~\cite{ki97,is00}.
In this case, the retardation effect is neglected by taking $q_0=0$ as
in many other calculations~\cite{se87,bu87,br90}. This makes the
calculations much easier. The small difference in proton and neutron
masses is retained in our calculations. To go beyond the lowest-order
ROM, we have also re-calculated all key quantities using the RPA.
Results of these calculations will be compared with those using the
lowest-order ROM.

\subsection{ Results with the lowest-order ROM}
\begin{figure}[tbh]
\begin{center}
\vspace*{-20mm}\includegraphics[width=0.7\textwidth]{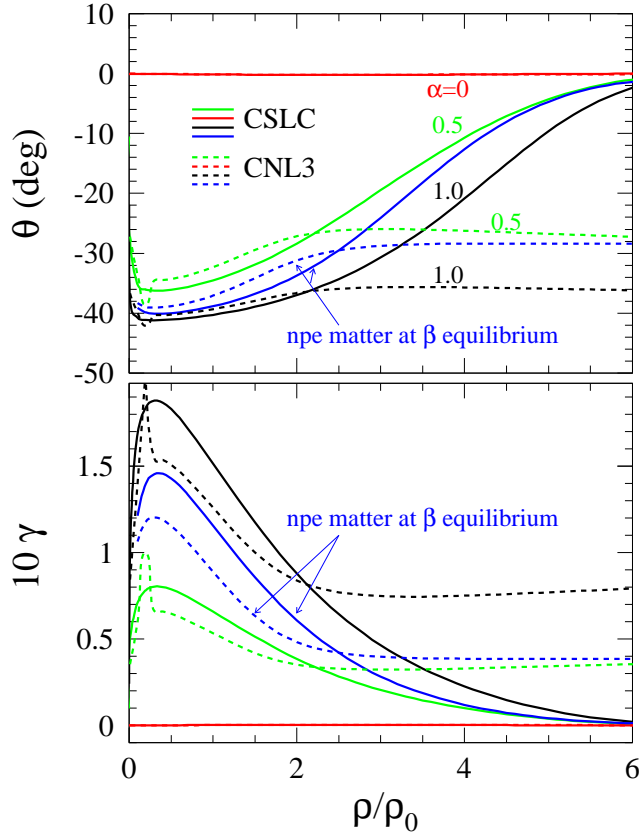}
\vspace*{-10mm}
\end{center}
\caption{(Color online) The mixing angle (the upper panel) and the parameter
$\gm$ (the lower panel) as a function of density for different isospin
asymmetries and npe matter at $\bt$ equilibrium with the CSLC and CNL3.
\label{figang}}
\end{figure}
The $\r-\om$ mixing angle is obtained by reproducing the total energy
density including the polarization diagram as we outlined in the
previous section. Fig.~\ref{figang} depicts the  $\r-\om$ mixing angle
$\theta$ and the dimensionless parameter $\gm$ for different isospin
asymmetries $\af=(\r_n-\r_p)/\r_B$ as a function of density with the
CSLC (solid lines) and CNL3 (dashed lines). The model CSLC is simply
the model SLC plus the in-medium CSB given by the ROM. Since the
in-medium CSB is negligible in symmetric matter, both the CSLC and SLC
have almost the same EOS for symmetric matter. In order to have the
same $E_{\rm sym}=31.6$ MeV at saturation density
$\r_0=0.16fm^{-3}$~\cite{ch05,li05} as the SLC, $G_\r$ (i.e., $G^*_\r$
at $\r=0$) in the CSLC is readjusted to 2.82.  Similarly, the model
CNL3 is constructed from the original model NL3 plus the in-medium CSB
but with $G_\r=3.02$ to have the same symmetry energy at saturation
density as given by the NL3. The $\theta$ is negative which is opposite
to that in Refs.~\cite{ki97,is00} where only the temporal component of
the polarization was considered. Our result is consistent with that in
Ref.~\cite{du97} where the $\theta$ was obtained from accounting
higher-order contributions in the Dyson equation.  In free space, the
$\theta$ goes to zero in our calculation because the observed energy
density, actually obtained by subtracting the vacuum expectation value,
is zero in the free space. The mixing angle is negligibly small for
$\af=0$, and increases with $\af$ as one expects. For pure neutron
matter, the magnitude of the $\theta$ reaches the maximum value of
about 40$^\circ$ at very low densities and then decreases with the
increasing density. Since neutron stars are among the most neutron-rich
and mysterious objects in the Universe, it is interesting to also
examine the mixing angle in the neutron-proton-electron (npe) matter at
$\bt$ equilibrium. As shown also in Fig.~\ref{figang}, the mixing angle
in the npe matter at $\bt$ equilibrium is generally between those for
pure neutron matter and $\af=0.5$. The $\gm$, shown in the lower panel
of Fig.~\ref{figang}, varies from almost zero in symmetric matter to
about 0.2 in highly isospin-asymmetric matter at low densities, while
at high densities it decreases with the CSLC and almost saturates with
the CNL3, similar to the mixing angle $\theta$.  We stress here that in
all cases considered, the mixing angle is negative with its magnitude
well below 45$^\circ$. This is rather different from that obtained in
Refs.~\cite{ki97,is00}. We can thus avoid naturally the unphysical
result that the $\r$-meson mass becomes imaginary when the magnitude of
$\theta$ is close to 45$^\circ$ as seen in Eq.(\ref{eq:mm}).

The nuclear EOS obtained with the SLC differs mainly in the high
density region from that with the NL3. The latter also has no nonlinear
terms for vector mesons and thus guarantees the simplicity of the
linear transformation in Eq.(\ref{trans}). It is thus interesting to
compare their predictions on the mixing angles. First of all, it is
seen from Fig.~\ref{figang} that the maximum magnitude of the mixing
angle with the CNL3 is also well below 45$^\circ$. Secondly, it is seen
that the difference between predictions with the CSLC and CNL3 is
mainly at high densities. While the CNL3 predicts approximately
constant and large mixing angles, the values with the CSLC become
almost zero at high densities. This is understandable since the mixing
angle is originated from the nucleon polarization. In the CSLC, the
vertices from the polarization are dictated by the BR scaling that
plays an important role in suppressing the medium-induced ROM
contribution as seen in the Eq.(\ref{edpot2}) and Fig.~\ref{fmix1}. In
the CNL3, however, such a suppressing factor does not exist. Therefore,
there is a large difference in the predicted mixing angle. The
difference for the parameter $\gm$ between the CSLC and CNL3, shown in
Fig.~\ref{figang}, can be understood similarly.

\begin{figure}[tbh]
\begin{center}
\vspace*{-20mm}\includegraphics[width=0.7\textwidth]{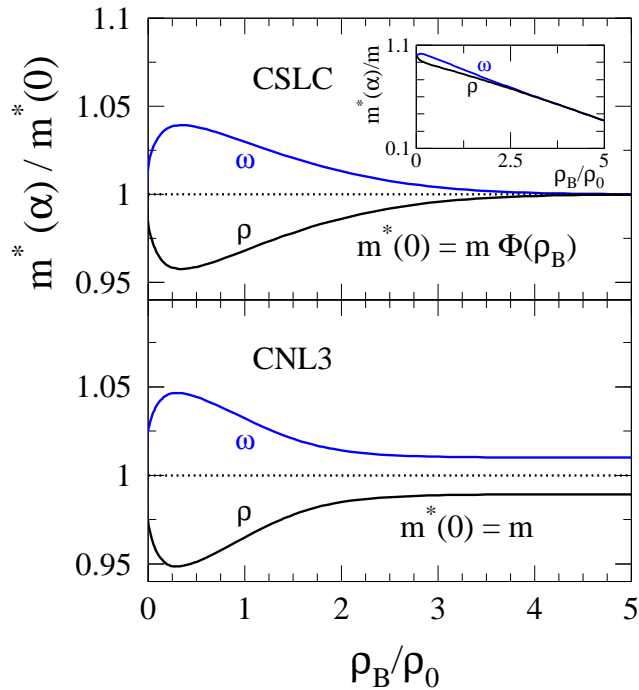}
\vspace*{-10mm} \end{center} \caption{(Color online) The $\r$ and
$\om$ meson masses in the npe matter at $\bt$ equilibrium scaled by
their respective masses in symmetric matter. The inset shows their
masses scaled by their respective masses in vacuum. $\Phi(\r_B)$ is
the scaling function of meson masses~\cite{ji07}. \label{figmm}}
\end{figure}

We now turn to the  effects of the medium-induced ROM on properties
of vector mesons in asymmetric matter. As an example, we consider the
npe matter at $\beta$ equilibrium. Fig.~\ref{figmm} displays the
changes of the vector meson masses versus density in the npe matter
at $\bt$ equilibrium for the CSLC (upper window) and CNL3 (lower
window) models. To separate effects due to the medium-induced ROM
from those due to the mass dropping scenario, we plot separately the
$\r$ and $\om$ meson masses scaled by their respective masses in
symmetric matter in the main frame, and the masses scaled by their
respective masses in vacuum in the CSLC case, which are in fact the
BR scaling functions for the meson masses~\cite{ji07}, in the inset
of the upper window. It is seen that the relatively large
modification occurs mainly within the low density region. Similar to
the high density behaviors of the mixing angle with the CSLC and
CNL3, for the same reason the two models predict slightly different
masses at high densities. With the CNL3, the splitting of the scaled
$\r$ and $\om$ masses stays almost unchanged at high densities,
whereas with the CSLC the splitting vanishes quickly. Nevertheless,
the modifications to the scaled masses are quite small and remain at
the level of about 5 percent or less for both models.

The mass dropping relative to its vacuum value is a manifestation of
the partial restoration of chiral symmetry, as suggested in many
effective QCD theories. It is worthwhile to mention that some evidences
for the dropping meson masses was found from studying the dilepton
spectra observed at the CERN SPS~\cite{li95,ag95} in the 90's.
Recently, a downward shift of the $\om$ meson mass was observed at the
KEK~\cite{kek} and the ELSA-Bonn~\cite{tap}. At saturation density, the
mass ratios ($m_i^*(\af)/m_i(0), i=\r,\om$) with the CSLC are predicted
in the npe matter at $\bt$ equilibrium ($\af=0.92$) to be 0.846 and 0.9
for the $\r$ and $\om$ mesons, respectively. At $\af=0.2$ which is
approximately the average isospin asymmetry for the $^{208}$Pb nucleus,
the scaled masses at saturation density are 0.878 and 0.87 for the $\r$
and $\om$ mesons, respectively. This is equivalent to a splitting of
their in-medium peaks by about 17 MeV.

\begin{figure}[tbh]
\begin{center}
\vspace*{-20mm}\includegraphics[width=0.7\textwidth]{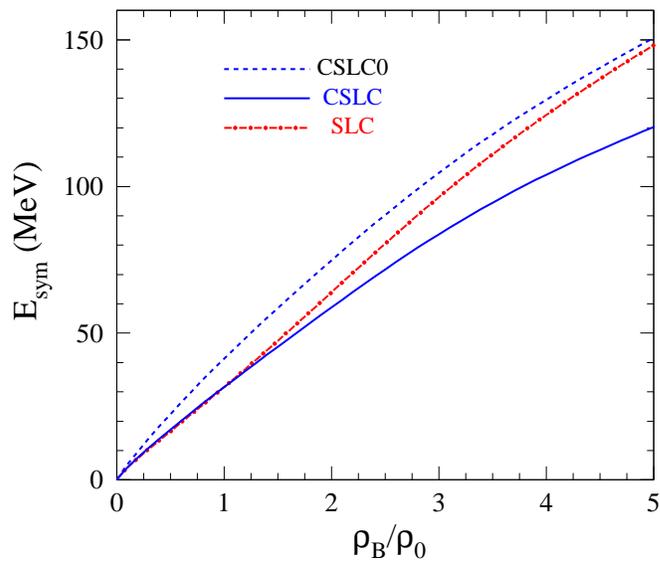}
\vspace*{-10mm} \end{center} \caption{(Color online) The symmetry
energy as a function of density for three cases (see text).
\label{figse}}
\end{figure}
To examine effects of the medium-induced ROM on the symmetry energy
defined as $E_{sym}(\rho)\equiv(\frac{\partial^2\mathcal{E}}{\partial
\alpha^{2}})_{\rho}/2\rho$, we compare in Fig.~\ref{figse} results for
three cases: the SLC, CSLC0 (SLC plus the additional energy density
with the medium-induced ROM), and the CSLC. Though this additional
energy density is negligible in symmetric matter, it becomes important
in isospin-asymmetric matter. Consequently, the symmetry energy
reflecting the cost to go away from symmetric matter is significantly
modified. It is seen that the effect on the symmetry energy is
significant at densities around $1-2\r_0$. At higher densities, the
modification fades away because the vector meson coupling constants
(see Eq.(\ref{edpot2})) tends to zero at high densities according to
the BR scaling~\cite{ji07}. To see more clearly the influence of the
medium-induced ROM on the density dependence of the symmetry energy, we
also compare results obtained with the SLC and CSLC. It is interesting
to see that a large softening effect is observed at high densities.
While considerable progress has been made recently in constraining the
density dependence of the symmetry energy at subsaturation densities
using data from heavy-ion reactions~\cite{ts04,ch05,li05}, experimental
information about the symmetry energy at supra-normal densities just
starts to emerge\cite{Fopi07,xiao08}. The effect of the medium-induced
ROM observed here is significant. It further adds to the importance to
determine experimentally the symmetry energy at supra-normal densities.
Hopefully, heavy-ion reactions induced by high energy radioactive beams
will make this possible in the near future~\cite{li02}.

Now, let's understand the in-medium CSB more clearly. Firstly,  the
effects of the in-medium CSB are related to the modification to the
symmetry energy. We may decompose the polarization of the ROM into two
parts: the density-dependent part and the Feynman (or, the vacuum)
part. The Feynman part changes the sign of the correction to the energy
functional as the charge-conjugate operator is acted on, and it would
result in a very small contribution to the energy term linear in
isospin asymmetry. This part is responsible for the usual CSB effect
within the ROM picture as one performs the charge-conjugate operation.
Here we do not see significant numerical contributions to the symmetry
energy. However, the density-dependent part does not change sign for
such an operation and mostly contribute to the symmetry energy term
quadratic in isospin asymmetry. Secondly, the meaning of the in-medium
CSB may concern the in-medium nucleon-nucleon interactions. In fact, if
one discusses the in-medium nucleon-nucleon scatterings, the in-medium
CSB has the content of the medium-induced ROM, and it is amplified
dynamically by the incomplete cancelation between the proton and
neutron loops in isospin-asymmetric matter.

Next, we discuss the causality issue in asymmetric nuclear matter. We
examine the pressure by taking into account the rearrangement term.
In addition to the contribution from the density dependence of the
meson masses and coupling constants induced by the BR scaling, the
rearrangement term in the CSLC includes several $\theta$-entangled
terms. In the CNL3, however, the rearrangement term  is only from a
single $\theta$-related source. In Refs.~\cite{ki97,is00}, a large
$\theta$ close to $45^\circ$ was obtained in isospin-asymmetric
matter. In this case, unfortunately, the neglected rearrangement term
actually dominates the pressure. More specifically, it gives the
leading difference between the pressure and the energy density, i.e.,
\begin{equation}\label{rrr1}
p-{\mathcal{E}}\ra\frac{m_0^{*2}-m_1^{*2}}{2{m_\r^*}^4}\frac{\sin
2\theta}{\cos^2 2\theta}(g_{n\om}^*\r_n-g_{p\om}^*\r_p)^2\frac{\pp \theta}{\pp
\r},
\end{equation}
where $m_0^*$ and $m_1^*$ are taken their respective vacuum values in
Refs.~\cite{ki97,is00}. It diverges to positive infinity for
$\theta\ra 45^\circ$ and $\pp \theta/\pp \r>0$~\cite{ki97,is00}. This
certainly violates the causality limit of $p\leq{\mathcal E}$. In
fact, Eq.(\ref{rrr1}) remains finite for physical cases. The explicit
violation of the causality indicates that the extremely large mixing
angle obtained in Refs.~\cite{ki97,is00} is clearly inappropriate. It
might be useful to comment here that the rearrangement term itself,
necessarily required by the thermodynamic consistency, allows us to
check the applicability or rationality of the methodology applied
under the extreme conditions of density and isospin asymmetry.  To be
more quantitative, in our treatment we examine the causality
condition in pure neutron matter by displaying the relation between
the energy density and the pressure in appendix \ref{appendb}.

\subsection{ Results within the RPA}
\begin{figure}[tbh]
\begin{center}
\vspace*{-20mm}\includegraphics[width=0.7\textwidth]{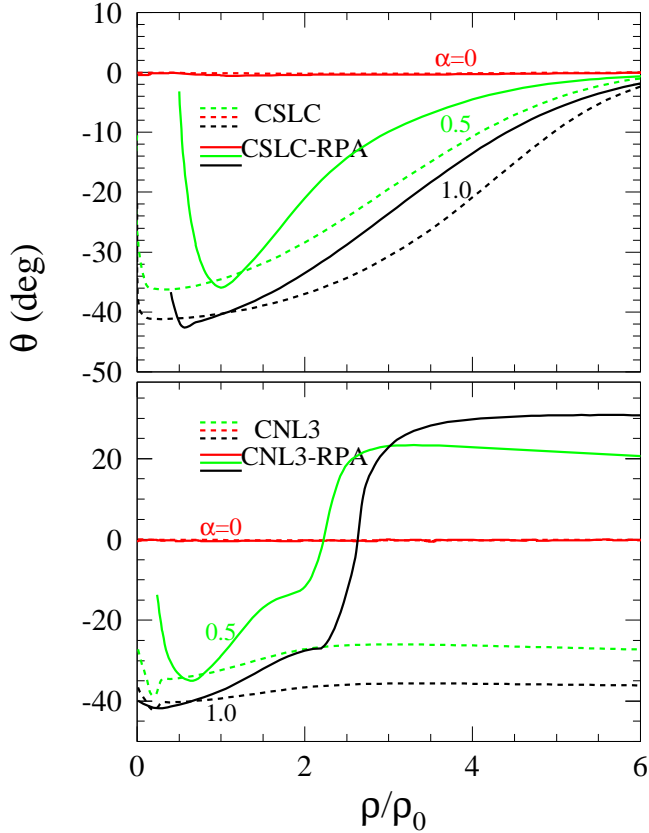}
\vspace*{-10mm}
\end{center}
\caption{(Color online) The mixing angle as a function of density for
different isospin asymmetries.  The upper panel is for the CSLC-RPA,
and the lower is for the CNL3-RPA. For comparison, the results
without the RPA correction are also exhibited.\label{angrpa}}
\end{figure}

The RPA equation contains both the real and imaginary parts of the
polarizations. As we can see, the imaginary part goes away completely
if the retardation effect is neglected, namely, by taking $q_0=0$.
Thus, to keep the imaginary parts of the polarizations in the RPA
equation, we have to account for the retardation effect. Here, we
define the CSLC-RPA0 as the model similar to the CSLC0 but with the ROM
calculated in the RPA. The model CSLC-RPA is the same as the CSLC-RPA0
but with $G_\rho=2.8$ to reproduce the same symmetry energy at
saturation density as the SLC. Similarly, we can define the models
CNL3-RPA0 and CNL3-RPA (with $G_\rho=3.36$) in reference to the CNL3
and NL3.

Fig.~\ref{angrpa} shows the mixing angle with the CSLC-RPA (the upper
panel) and CNL3-RPA (the lower panel) for different isospin
asymmetries. It is seen that the absolute value of the mixing angle
with the CSLC-RPA reaches its maximum around the saturation density.
Interestingly, this maximum is very close to the one obtained from
the lowest-order ROM. Away from the maximum point, the magnitude of
the mixing angle decreases compared to that from the CSLC. This
indicates that the high-order contributions included in the RPA
cancel partially the lowest-order ROM. Concretely, this partial
cancellation occurs mainly between the longitudinal modes with and
without the RPA correction since both the transverse mode of the
lowest-order ROM and its RPA correction are small in the whole
density region. It is known that the transverse mode of the
lowest-order polarization is small compared to the longitudinal one.
In the SLC model, the small value of the transverse RPA correction is
mainly due to the following two facts. Firstly, the small transverse
mode of the lowest-order ROM does not increase appreciably with the
density because of the suppression added by the BR scaling. Secondly,
the transverse eigencondition $\ep_T=0$ can not be satisfied
accordingly. However, the RPA correction in the NL3 model gives rise
to a very different feature at high densities. As shown in the lower
panel of Fig.~\ref{angrpa},  the mixing angle with the CNL3-RPA
starts increasing from about twice the normal density and then
becomes positive at density $\r\ge2.5\r_0$. Similar to the case with
the CSLC-RPA, the RPA correction to the longitudinal mode reaches its
maximum in the vicinity of the saturation density and goes steadily
without large changes at high densities in the CNL3-RPA calculations.
Thus, this very different behavior from that obtained with the
CSLC-RPA can only be due to the distinct property of the RPA
correction to the transverse mode of the ROM with the CNL3-RPA.
Within the CNL3-RPA, the small transverse mode of the lowest-order
polarization increases with the density and the transverse
eigencondition $\ep_T=0$ can be satisfied at high densities
($\r\geq2.0\r_0$).  This then results in a large RPA correction to
the transverse mode of the ROM, responsible for the sign change of
the mixing angle. Similarly, at high densities the large difference
between the CNL3 and CNL3-RPA results can be explained mainly by
using the transverse mode property of the RPA polarization in the
CNL3-RPA. Next, we examine the mixing angle in the low-density
region. At very low densities ($\r\le0.3\rho_0)$, the mixing angle in
highly isospin asymmetric matter can be shifted to positive values
with both the CSLC-RPA and CNL3-RPA models. This shift is due to the
reduced Pauli blocking on the various intermediate excitations
(polarizations) at very low densities.

\begin{figure}[tbh]
\begin{center}
\vspace*{-20mm}\includegraphics[width=0.7\textwidth]{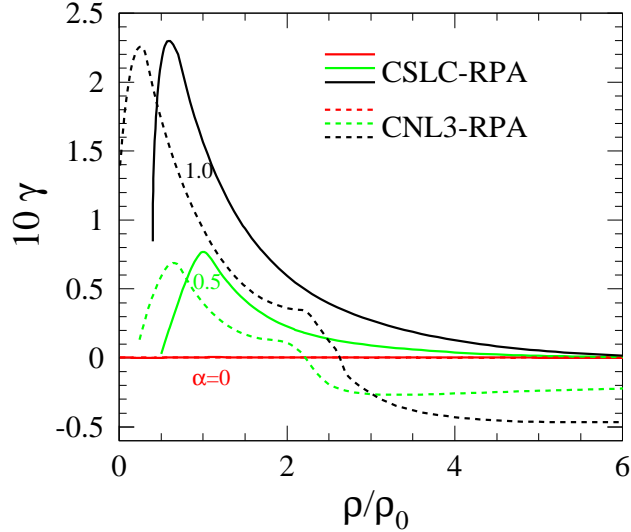}
\vspace*{-10mm}
\end{center}
\caption{(Color online)  The parameter $\gm$ with the CSLC-RPA and CNL3-RPA for
various isospin asymmetries.  \label{figgm}}
\end{figure}

Fig.~\ref{figgm} depicts the parameter $\gm$ obtained with the CSLC-RPA
and CNL3-RPA. Compared to the results shown in Fig.~\ref{figang}, the
RPA correction shifts both the maximum value and its position
moderately in the low-density region for both models. At high
densities, larger effects of the RPA correction start to appear around
$2\r_0$. It is seen that the difference in $\gm$ between the CSLC-RPA
and CNL3-RPA calculations also becomes more appreciable at high
densities ($\r\geq 2\r_0$). These observations are consistent with
those from studying the mixing angle shown in Fig.~\ref{angrpa}.

\begin{figure}[tbh]
\begin{center}
\vspace*{-20mm}\includegraphics[width=0.7\textwidth]{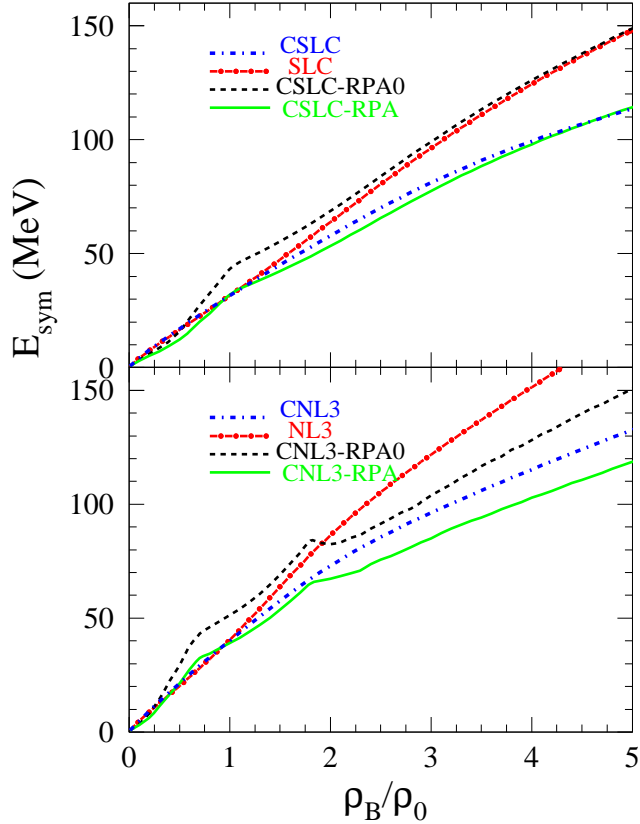}
\vspace*{-10mm} \end{center} \caption{(Color online) The symmetry energy as a
function of density including the RPA corrections within the CSLC (upper
panel) and CNL3 (lower panel). Here,  the retardation effects are also
included in the calculation within the CSLC and CNL3. \label{figpir}}
\end{figure}

Next, we turn to the RPA correction to the symmetry energy. Shown in
Fig.~\ref{figpir} is a comparison of the symmetry energies obtained
with the ROM with or without the RPA correction. We first examine in
the upper panel effects of the ROM within the RPA on the symmetry
energy with the CSLC-RPA0 and CSLC-RPA calculations. In the SLC model,
consistent with the analysis for the mixing angle in the above, the RPA
correction to the energy density is dominantly from the longitudinal
mode. Comparing the results obtained with the CSLC-RPA0 and SLC, we can
see that with the increasing density the medium-induced ROM with the
RPA affects only slightly the symmetry energy. This is consistent with
the effects on the mixing angle shown in Fig.~\ref{angrpa}. Moreover,
by comparing the symmetry energy with the CSLC-RPA0 to that with the
CSLC0, see Fig.~\ref{figse}, we see that the RPA correction to the
lowest-order ROM is small at high densities. Since the RPA correction
reaches its minimum at saturation density as shown earlier in
Fig.~\ref{angrpa}, the bump is interestingly formed in the CSLC-RPA0
results. Noticing also that the RPA correction is almost negligible at
saturation density, we can infer that the medium-induced ROM with the
RPA modifies the symmetry energy as much as the lowest-order ROM at
saturation density.  When the symmetry energy at saturation density is
constrained with the empirical value, it  is then considerably softened
at high densities. This softening is clearly seen in Fig.~\ref{figpir}
by comparing the results obtained with the CSLC-RPA and SLC. At high
densities, there is almost no difference between the symmetry energies
obtained with the CSLC and CSLC-RPA. Furthermore, at subsaturation
densities the symmetry energy with the CSLC-RPA is also softened with
respect to the SLC due to the high-order correlations included in the
RPA calculations.

In the lower panel of Fig.\ref{figpir}, we compare results obtained
with or without the RPA correction based on the NL3 parameter set. The
CNL3-RPA0 and CNL3-RPA results display some noticeable bumpy
structures. The bump formed at the lower density is associated
dominantly with the RPA correction to the longitudinal mode of the ROM,
since the transverse RPA correction is small at low densities. Thus, it
can depend sensitively on the nonlinear self-interaction of the $\sg$
meson through the unperturbed propagator $\Dt$, (see
Eqs.(\ref{rpa4}),(\ref{eq:el})). In the nonlinear RMF model (such as
the NL3), the $\sg$-meson mass in the unperturbed propagator $\Dt$
becomes the effective mass due to its nonlinear
self-interaction~\cite{re99}. Once the self-interaction of the $\sg$
meson is neglected in the $\Dt$, we find that the bump at the lower
density is then shifted upwards and centralized around the saturation
density, similar to the case with the CSLC-RPA0 which is free of the
nonlinear meson self-interaction. As the nonlinear self-interaction of
the $\sg$-meson is necessarily included in the $\Dt$ in calculations
with the NL3 model, the symmetry energy at subsaturation density is
expected to be modified differently by the RPA compared to calculations
with the SLC model. Noticeably, starting about $2\rho_0$ the symmetry
energy becomes softer in calculations with the RPA. Consequently, in
the high-density region, the difference between the NL3 and CNL3-RPA0
results (lower panel) is significantly larger than that between the SLC
and CSLC-RPA0 results (upper panel). This feature is consistent with
that shown for the mixing angle and the parameter $\gm$ in
Fig.~\ref{angrpa} and ~\ref{figgm}, respectively. Again, this can be
attributed to the different properties of the RPA correction to the
transverse mode of the ROM in the CSLC-RPA0 and CNL3-RPA0 calculations.
Moreover, the lowest-order ROM can largely soften the symmetry energy
as seen clearly by comparing the results obtained with the NL3 and
CNL3. This softening by the lowest-order ROM is dictated by the
longitudinal mode, while the transverse mode plays only a minor role.
We note that the RPA correction to the longitudinal ROM suppresses
considerably the softening due to the lowest-order (longitudinal) ROM.
Thus, it is worth stressing that in the nonlinear RMF model (NL3) the
RPA correction to the transverse mode of the ROM plays a crucial role
in softening the symmetry energy at high densities. Comparing the
symmetry energy with the CNL3-RPA to that with the NL3, it is seen that
the softening is very significant at high densities. Interestingly, the
symmetry energy at high densities with the CNL3-RPA is comparable to
that with the CSLC-RPA which features a typically soft symmetry energy
at high densities.

In this work, we have only considered the medium-induced ROM and its
impact on the vector meson properties and the symmetry energy. The
in-medium $\sg-\r$ mixing may also affect the symmetry energy. However,
its effect is expected to be much smaller than the ROM due to the small
$\sg-\r$ polarization in the low-momentum region~\cite{ro08}.
Nevertheless, a thorough investigation including the $\sg-\r$ mixing
may be an interesting topic for a future work. It is also worth noting
that in solving the RPA equation the Feynman parts of the polarizations
except for the ROM are neglected. The neglected Feynman parts may be
useful for studying the vacuum effect on the symmetry energy. This of
course needs a lot more efforts in renormalized models.  Finally, we
also mention that the in-medium CSB effect on the EOS of symmetric
nuclear matter is negligible. This finding is consistent with earlier
conclusions as we discussed in the introduction.

\section{Summary}
\label{summary}

In summary, using the RMF models with and without the BR scaling at
high densities we studied the medium-induced ROM and its effects on the
in-medium masses of the $\rho$ and $\omega$ mesons and the density
dependence of the symmetry energy  in isospin-asymmetric dense nuclear
matter. The mixing angle is obtained by reproducing the additional
energy density from the $\omega-\rho$ conversion through the
polarization diagram. We took into account the rearrangement term.
Significantly different predictions from some earlier studies are made
especially at high densities. We found that the symmetry energy is
significantly softened instead of being stiffened at high densities by
the medium-induced ROM. The $\r-\om$ mass splitting due to the
medium-induced ROM in highly neutron-rich matter is appreciable mostly
at low densities and is small at high densities. While for the RMF
model with the BR scaling, the $\r-\om$ mass splitting  further
vanishes at high densities. No large drop of the $\r$ meson mass is
observed in asymmetric matter, contrary to some earlier studies by
others.  The medium-induced ROM is also studied with the RPA. The
symmetry energy is significantly softened by the ROM in the RPA at both
supra-saturation and sub-saturation densities. We also find that such a
softening at sub-saturation densities due to the $\om-\r$ mixing can be
partly suppressed by the nonlinear self-interaction of the $\sg$ meson.

\section*{Acknowledgement}
We thank S. A. Chin, Lie-Wen Chen, and P. Krastev for helpful discussions.
The work was supported in part by the US National Science Foundation Grants
PHY-0652548 and PHY-0757839, the Research Corporation under Grant No. 7123
and the Texas Coordinating Board of Higher Education Grant No.
003565-0004-2007, the National Natural Science Foundation of China under
Grant Nos. 10975033, 10575071 and 10675082, the China Jiangsu Provincial
Natural Science Foundation under Grant No. BK2009261, the Knowledge
Innovation Project of the Chinese Academy of Sciences under Grant No.
KJXC3-SYW-N2, and the China Major State Basic Research Development Program
under Contract No. 2007CB815004.

\appendix
\section{Dielectric functions}
\label{appenda}

Prior to giving the dielectric functions used in Eq.(\ref{rpa4}), we define the
nucleon polarizations according to the Feynman rules~\cite{ch77,se86}:
\begin{equation}\label{pdef}
\Pi^{ab}(q)=-i\int_{-\infty}^\infty \frac{d^4k}{{(2\pi)}^4} {\rm
Tr}[iG(k)\Gm_a iG(k+q)\Gm_b],
\end{equation}
where the  vertices $\Gm_a=ig_\sg^*,-iG_\om^*\gm_\mu$, and
$-iG_\r^*\tau\gm_\mu$ for $a=\sg,\om$, and $\rho$, respectively. For
the neutral $\rho$ meson, the vertex is $-iG_\r^*\tau_3\gm_\mu$. For
the nucleon propagator, we use the expression in Ref.~\cite{se86}. With
the general definition of polarizations, one can give the expression of
the $\Pi_{\mu\nu}^{\rho\om}$ in Eq.(\ref{eq:pola2}). The elements of
the polarization matrices in Eq.(\ref{rpa3}) can also be written out
explicitly according to the above definition.

The nucleon polarization for the vector meson can be appropriately
decomposed into longitudinal and transverse modes such as in
Eq.(\ref{eq:pola1}), (\ref{rpa2}) and (\ref{rpa3}). In the RPA equation
for the nucleon polarization for the vector meson, the longitudinal and
transverse dielectric functions in Eq.(\ref{rpa4}) are given as:
\begin{eqnarray}\label{eq:el}
  \epsilon_L &=& (1-\Dt\Pi^{\sg\sg})(1-D_\om\Pi^{\om\om}_L) (1-D_\r\Pi^{\r\r}_L)
  -D_\om D_\r(1-\Dt\Pi^{\sg\sg})(\Pi_L^{\r\om})^2\nnb\\
   & &-\Dt D_\om (1-D_\r\Pi^{\r\r}_L)\frac{q^2}{\bar{q}^2}(\Pi_0^{\sg\om})^2 -
 \Dt D_\r (1-D_\om\Pi^{\om\om}_L)\frac{q^2}{\bar{q}^2}(\Pi_0^{\sg\r})^2\nnb\\
 & &-2\Dt D_\om D_\r\frac{q^2}{\bar{q}^2}
\Pi^{\sg\om}_0\Pi_0^{\sg\r}\Pi_L^{\r\om},
\end{eqnarray}
and
\begin{equation}\label{eq:et}
   \epsilon_T = (1-D_\om\Pi_T^{\om\om}) (1-D_\r\Pi^{\r\r}_T)
  -D_\om D_\r(\Pi_T^{\r\om})^2.
\end{equation}
For simplicity, in the text we do not give the explicit expressions for
all other polarizations except for the real part of the $\Pi^{\rho\om}$
given since they can be found easily in literatures, e.g., see
Refs.~\cite{ch77,ho91,lim89}.

\section{Causality condition in pure neutron matter at high densities}
\label{appendb}

\begin{figure}[tbh]
\begin{center}
\vspace*{-20mm}\includegraphics[width=0.7\textwidth]{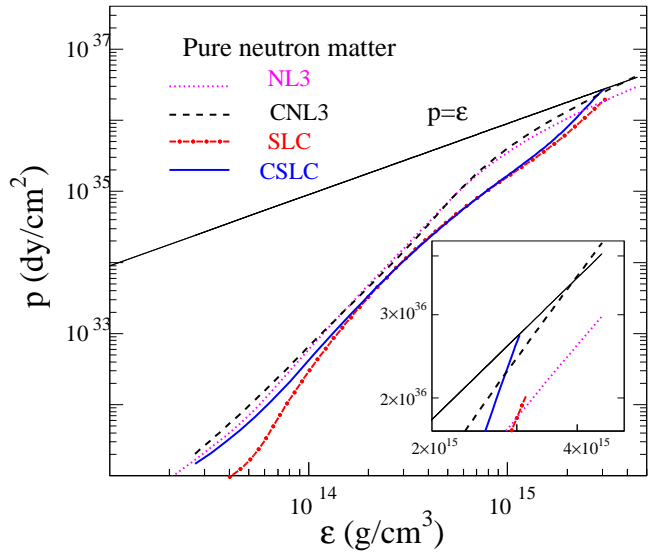}
\vspace*{-10mm}
\end{center} \caption{(Color online) The pressure versus energy density in pure
neutron matter. The inset shows the pressure within the small
interval.\label{figprss}}
\end{figure}

In order to examine the causality condition in our calculations,
we display in Fig.~\ref{figprss} the pressure versus energy
density for pure neutron matter. Comparing the results obtained
using the CSLC and SLC, one sees that the in-medium CSB increases
the pressure appreciably in both the low and high density regions.
The same conclusion but to a lesser extent can be drawn from
comparing results obtained using the CNL3 and NL3. Unfortunately,
our numerical calculations, as illustrated in the inset for the
high density tail, indicate that the causal condition
$p\leq{\mathcal E}$ is broken with the CNL3 for $\r\geq7.2\r_0$
(here, $r_0=0.16$fm$^{-3}$). These features displayed in
Fig.~\ref{figprss} can be elaborated by virtue of the
rearrangement term $-\Sg_0~\r$ in Eq.(\ref{eqp1}). The $\Sg_0$ is
given by
\begin{equation}\label{rearr1}
\Sg_0=\Sg_{0\theta}+\Sg_{0BR},
\end{equation}
where
\begin{equation}\label{rearra}
  \Sg_{0\theta} = -\frac{\pp {\mathcal E}}{\pp\theta}
\frac{\pp\theta}{\pp\r}, ~~~
  \Sg_{0BR} =-\sum_{i}\frac{\pp {\mathcal E}}{\pp\phi_i}
\frac{\pp\phi_i}{\pp\r},\nnb
\end{equation}
with $\phi_i$ being the various scaling functions used in the SLC
model~\cite{ji07}. For the CSLC calculations, since the $\theta$
approaches zero at very high densities, the $\Sg_{0BR}$ is little
changed by the ROM. Thus, the increase of the pressure at very high
densities with the CSLC compared to the SLC is due to the term
$\Sg_{0\theta}$. In pure neutron matter, the factor ${\pp {\mathcal
E}}/{\pp\theta}$ in the $\Sg_{0\theta}$ can be derived as
\begin{equation}\label{rearr2}
\frac{\pp {\mathcal E}}{\pp\theta}=-
\half\frac{(m_0^{*2}-m_1^{*2})\sin 2\theta}{\cos^2
2\theta}(\frac{g_{n\om}^{*2}}{m_\om^{*4}}-
\frac{g_{n\r}^{*2}}{m_\r^{*4}})\r_n^2 +(\frac{1}{m_\om^{*2}}-
\frac{1}{m_\r^{*2}})g^*_{n\om}g^*_{n\r}\r_n^2.
\end{equation}
In the CSLC, the first term dominates at very high densities because
the BR scaling renders the effective masses of the vector mesons to
be close to zero. With $\pp\theta/\pp\r>0$ and $\theta\ra-0$ (see
Fig.~\ref{figang}), the rearrangement term $-\Sg_{0\theta}~\r$ is
positive, leading to the increase of the pressure at very high
densities. On the other hand, since the masses of vector mesons are
very large in the CNL3 compared to the CSLC, the violation of the
causality limit at very high densities is dictated by the second
term.  We note here that the first and second terms of
Eq.(\ref{rearr2}) have opposite signs for negative $\theta$ values.
While in the CNL3 calculations, unlike the CSLC case, one has
$\pp\theta/\pp\r<0$ at very high densities. Thus, the second term of
Eq.(\ref{rearr2}) can increase the pressure and further violate the
causality limit in the CNL3. It is useful to note that the reason for
the causal breakdown here is different from the one in
Refs.~\cite{ki97,is00}. In the latter, the inappropriate inducement
produces a very large mixing angle close to 45$^\circ$ that is much
larger than the one obtained here with the CNL3. Here, the violation
of the causality limit is actually due to an inconsistent treatment.
Considering the whole set of coupled equations for the propagators
and self-energies, the Lorentz covariance is respected only
approximately since we neglected the ROM loop-diagram correction to
the self-energies. Consequently, this may lead to a formal violation
of the Hugenholtz-Van Hove (HVH) theorem which is a conservation law
ensuring the thermodynamical consistency. In isospin-asymmetric
matter, the HVH theorem can be written as
\begin{equation}\label{eqhv}
\frac{{\mathcal E}}{\r}+\r\frac{\pp({\mathcal E}/\r)}{\pp\r}=
\half[\mu_n(1+\af)+\mu_p(1-\af)],
\end{equation}
where $\mu_n$ and $\mu_p$ are the neutron and proton chemical
potentials, respectively.  Since the nucleon chemical potentials
are determined by the nucleon effective Fermi energies and
self-energies, the neglect of the loop-diagram correction to the
self-energies in evaluating the total energy density from the
lagrangian (\ref{eq:lag1}) results in a formal violation of the
HVH theorem. It is certainly a drawback of the CNL3 calculations.
Nevertheless, as shown in Fig.~\ref{figprss}, this violation is
very weak and only visible at very high densities. Also, we notice
that this situation is not very rare in calculations using models
going beyond the mean field level. For instance, it is widely
acknowledged that approaches such as the relativistic and
non-relativistic Brueckner theories also unfortunately encounter
similar problems in fulfilling accurately the HVH
theorem~\cite{ka80,len96,zuo99}.

In principle, a self-consistent treatment at least in the
relativistic Hartree-Fock framework is needed to resolve the
causality violation at very high densities. However, this is beyond
the scope of the present work. Here, the loop-diagram correction to
the self-energy is estimated to be tiny (roughly 2\% in pure neutron
matter), and the slight causal violation occurs only in highly
isospin-asymmetric matter at very high densities. Therefore, the
symmetry energy obtained in the vicinity of symmetric matter is not
expected to be affected appreciably. Compared to the CNL3
calculations, the causality preservation with the CSLC is due to the
fact that the mixing angle is almost zero at very high densities
because of the BR scaling used. As shown in Fig.~\ref{figprss}, the
CSLC curve ends at the critical density for the chiral
restoration~\cite{ji07}.

The features of Fig.~\ref{figprss} in the intermediate and low
density regions can also be understood. In the intermediate
density region, one has $\pp\theta/\pp\r>0$ within both the CNL3
and CSLC calculations. Thus the contributions from the first and
second terms in Eq.(\ref{rearr2}) cancel out to a large extent due
to their opposite signs. As shown in Fig.~\ref{figprss}, in the
intermediate density region, the pressure is little modified by
the $\theta$-relevant rearrangement term. At low densities,
however, a very distinct feature is exhibited by comparing the
difference between the pressures obtained with the CSLC and SLC to
that between those obtained with the CNL3 and NL3. In the SLC
calculations, the pressure is increased significantly by the ROM,
while it is just moderately increased in the NL3 calculations.
Since their mixing angles are very close in this density region,
the modification from the $\theta$-relevant rearrangement term
$\Sg_{0\theta}$ is comparable for both the SLC and NL3
calculations. Thus, this large difference can only be due to the
rearrangement term $-\Sg_{0BR}~\r$ appearing only in the SLC
parameter set.

\end{document}